\documentclass[amsmath, amssymb,10pt,aps,prb,twocolumn,notitlepage,showpacs,superscriptaddress]{revtex4-1}

\usepackage{graphicx}
\usepackage{calc}
\usepackage{braket}
\usepackage{ulem}   
\usepackage{slashed}
\usepackage{wasysym}
\usepackage{dsfont}
\usepackage{amsthm,amsmath,amsfonts,amssymb,verbatim,color}
\usepackage{graphicx}
\usepackage{subfigure}
\usepackage{bm}
\usepackage{epsfig,slashed}
\usepackage[T1]{fontenc}
\usepackage[colorlinks=true,citecolor=blue,linkcolor=blue,urlcolor=blue]{hyperref}
\newcommand{\be}{\begin{equation}}
\newcommand{\ee}{\end{equation}}

\begin{document}
\title{Logarithmic singularities and quantum oscillations in
 magnetically doped topological insulators}

\author{D. Nandi}
\affiliation{Department of Physics, Harvard University, Cambridge, MA 02138, USA}
\author{Inti Sodemann}
\affiliation{Max-Planck Institute for the Physics of Complex Systems, D-01187 Dresden, Germany}
\affiliation{Department of Physics, Massachussetts Institute of Technology, Cambridge, MA 02139, USA}
\author{K. Shain}
\affiliation{Department of Physics, Harvard University, Cambridge, MA 02138, USA}
\author{G.H. Lee}
\affiliation{Department of Physics, Harvard University, Cambridge, MA 02138, USA}
\affiliation{Department of Physics, Pohang University of Science and Technology, Pohang 790-784, Republic of Korea}
\author{K.-F. Huang}
\affiliation{Department of Physics, Harvard University, Cambridge, MA 02138, USA}
\author{Cui-Zu Chang}
\affiliation{Francis Bitter Magnet Lab, Massachusetts Institute of Technology, Cambridge, MA 02139, USA}
\affiliation{The Center for Nanoscale Science and Department of Physics, The Pennsylvania State
University, University Park, PA 16802-6300, USA }
\author{Yunbo Ou}
\affiliation{Francis Bitter Magnet Lab, Massachusetts Institute of Technology, Cambridge, MA 02139, USA}
\author{S.P. Lee}
\affiliation{Department of Physics, University of Alberta, Edmonton, Alberta T6G 2E1, Canada}
\affiliation{Department of Physics and Astronomy, The John Hopkins University, Baltimore, Maryland 21218, USA }
\author{J. Ward}
\affiliation{Department of Physics, Harvard University, Cambridge, MA 02138, USA}
\author{J.S. Moodera}
\affiliation{Francis Bitter Magnet Lab, Massachusetts Institute of Technology, Cambridge, MA 02139, USA}
\author{P. Kim}
\affiliation{Department of Physics, Harvard University, Cambridge, MA 02138, USA}
\author{A. Yacoby}
\affiliation{Department of Physics, Harvard University, Cambridge, MA 02138, USA}
\email{yacoby@physics.harvard.edu}
\date{\today}

\begin{abstract}
We report magnetotransport measurements on  magnetically doped (Bi,Sb)$_2$Te$_3$ films grown by molecular beam epitaxy. In Hallbar devices, we observe logarithmic dependence of transport coefficients in temperature and bias voltage which can be understood to arise from electron - electron interaction corrections to the conductivity and self-heating. Submicron scale devices exhibit intriguing quantum oscillations at high magnetic fields with dependence on bias voltage. The observed quantum oscillations can be attributed to bulk and surface transport.
\end{abstract}

\maketitle

\section{INTRODUCTION}

Breaking of time reversal symmetry in topological insulators can unlock exotic phenomenon such as the quantized anomalous Hall effect \cite{yu2010quantized,chang2013experimental,chang2015high,mogi2015magnetic,chang2015zero}, giant magneto-optical Kerr and Faraday effects \cite{tse2010giant}, the inverse spin-galvanic effect \cite{garate2010inverse}, image magnetic monopole effect \cite{qi2009inducing} and chiral Majorana modes \cite{qi2010chiral,he2017chiral}. Angle resolved photoemission spectroscopy measurements have revealed presence of a magnetic gap at the Dirac point as well as hegdehog spin texture in magnetic topological insulators \cite{chen2010massive,xu2012hedgehog}. Proximity coupling to a magnetic insulator such as EuS, yttrium iron garnet, and thulium iron garnet\cite{jiang2015independent,katmis2016high,tang2017above} or introducing magnetic dopants like Mn, Cr and V \cite{hor2010development,chang2013experimental,chang2015high} can remove time reversal symmetry. Such efforts have induced long range ferromagnetic order in topological insulators.

In this paper we explore magnetotransport in magnetically doped ultrathin films of (Bi,Sb)$_2$Te$_3$ to understand the role of different scattering mechanisms.  By studying the effect of temperature and voltage bias on the longitudinal and anomalous Hall resistances, we observe logarithmic dependences on temperature and voltage bias. Joule heating due to voltage bias increases the effective temperature of the hot electrons \cite{viljas2010electron}. The logarithmic singularities are originating from interplay of electron-electron interaction and disorder. We find that our observed logarithmic corrections quantitatively agree with
the Alshuler-Aranov theory of electron-electron interactions.  Furthermore, in submicron sized mesocale devices we observe quantum
oscillations that depend on voltage bias and weaken with increasing sample width.

\section{MATERIALS AND METHODS}

\begin{figure}[h!]
\begin{centering}
\includegraphics[width=8.5cm]{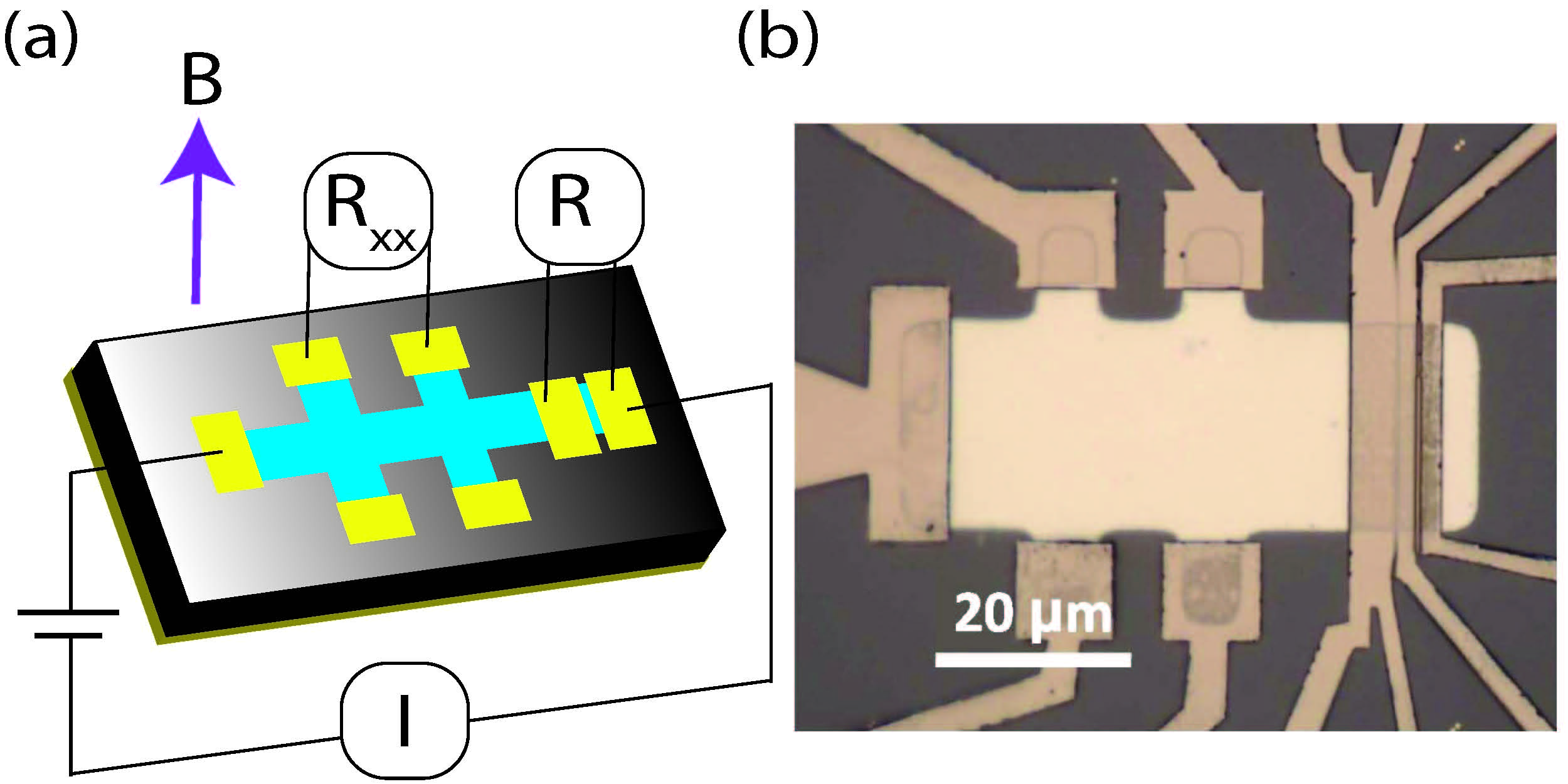}
\end{centering}
\caption{(color online) (a) Schematics of a magnetic topological insulator device. (b) Optical image of a Hallbar device H$_1$ and a two terminal device D$_1$ simultaneously patterned on magnetically (Vanadium) doped topological insulators.}
\label{device}
\end{figure}

The 4 quintuple layers (QLs) thick pristine and V-doped (Bi,Sb)$_2$Te$_3$ films are grown on SrTiO$_3$ (111) substrate by molecular beam epitaxy (MBE). The growth process was monitored in-situ by refection high-energy electron diffraction (RHEED) to ensure high quality films \cite{chang2015high, liu2016large, samarth2017quantum}. To prevent oxidation, a capping layer of 10 nm tellurium was deposited. The devices were fabricated employing standard photolithography and electron-beam lithography techniques.  The device schematic and optical image of a Ti/Nb/NbN contacted magnetic topological insulator film is shown in  Figs. \ref{device}(a)-\ref{device}(b). The transport measurements were done in a $^3$He-$^4$He dilution refrigerator using standard ac lock-in measurement techniques. Here we summarize results from a pristine  Hallbar device H$_0$, a V-doped  Hallbar device H$_1$ and a V-doped submicron scale device D$_1$.

\section{RESULTS AND DISCUSSION}

\begin{figure}[h!]
\begin{centering}
\includegraphics[width=8.5cm]{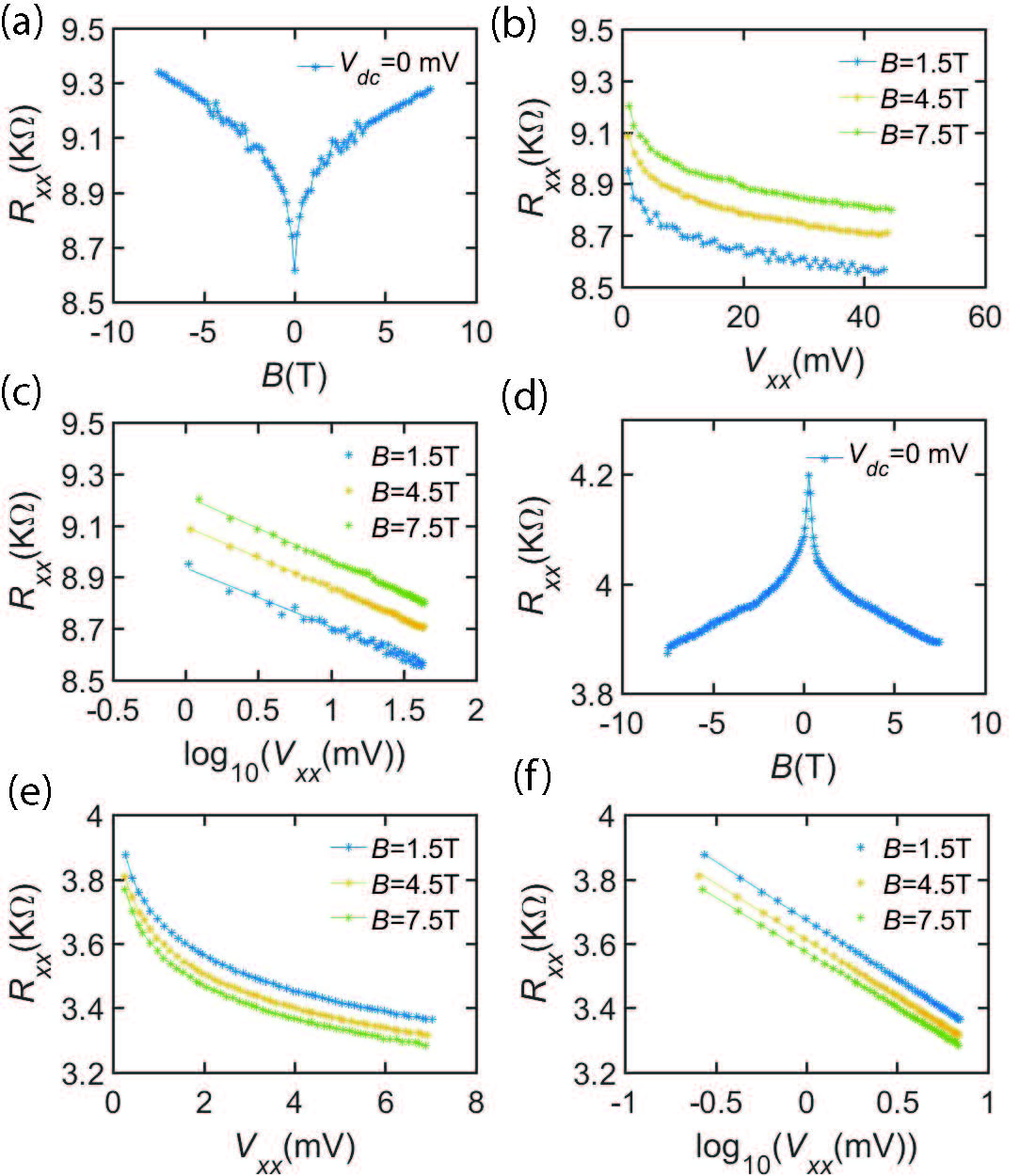}
\end{centering}
\caption{ Comparison of magnetotransport in pristine and V- doped 4 QL thick (Bi,Sb)$_2$Te$_3$ films. (a) Magnetoresistance dip at $B$=0 T in the sweep range $B$ = -7.5 to +7.5 T for pristine  (Bi,Sb)$_2$Te$_3$ samples. (b) Bias dependence of pristine (Bi,Sb)$_2$Te$_3$ film at $B$ = 1.5, 4.5 and 7.5 T. (c)  Logarithmic plot for bias dependence of 4 QL pristine (Bi,Sb)$_2$Te$_3$ film at $B$ = 1.5, 4.5 and 7.5 T. (d)  Magnetoresistance peak at $B$=0 T in the sweep range $B$ = -7.5 to +7.5 T for V-doped  (Bi,Sb)$_2$Te$_3$. (e) Bias dependence of  V-doped (Bi,Sb)$_2$Te$_3$ film at $B$ = 1.5, 4.5 and 7.5 T. (f) Logarithmic plot for bias dependence of V-doped (Bi,Sb)$_2$Te$_3$ film at $B$ = 1.5, 4.5 and 7.5 T.}
\label{longitudinal resistance}
\end{figure}

Magnetotransport measurements in a Hallbar device H$_0$ on pristine 4QL (Bi,Sb)$_2$Te$_3$ films presented in Fig. \ref{longitudinal resistance}(a) exhibit a dip in the longitudinal resistance $R_{xx}$  at $ B$ = 0 T which is attributed to weak antilocalization effect \cite{chen2010gate,chiu2013weak}. This is because in the presence of strong spin-orbit coupling time reversed trajectories have opposite spin orientations which lead to a destructive interference and a  resistance minimum \cite{hikami1980spin}.

We measured the bias dependence of longitudinal resistance $R_{xx}$ in the same device. The results are shown in Fig.  \ref{longitudinal resistance}(b) - Fig.  \ref{longitudinal resistance}(c) for a few different magnetic fields and exhibit a logarithmic dependence on voltage bias. Weak antilocalization is in itself a possible cause of logarithmic correction. However lowering temperature or voltage bias is expected to make weak antilocalization effect more pronounced thereby decreasing resistivity which is inconsistent with Fig. \ref{longitudinal resistance}(b) - Fig.  \ref{longitudinal resistance}(c).

Further by introducing magnetic impurities, the weak antilocalization effects can be heavily suppressed as has been reported in Fe-doped Bi$_2$Te$_3$ and Cr-doped Bi$_2$Se$_3$ films \cite{he2011impurity,liu2012crossover,lu2011competition}. Fig. \ref{longitudinal resistance}(d) shows the magnetoresistance in a V-doped (Bi,Sb)$_2$Te$_3$ has a peak  instead of a dip at $B$ = 0 T seen in pristine samples.  Even when the weak antilocalization effects are suppressed, the  longitudinal resistance  $R_{xx}$ in V-doped (Bi,Sb)$_2$Te$_3$  has a logarithmic dependence on voltage bias as shown in Fig. \ref{longitudinal resistance}(e) - Fig. \ref{longitudinal resistance}(f) at different magnetic fields.

\begin{figure}[h!]
\begin{centering}
\includegraphics[width=8.5cm]{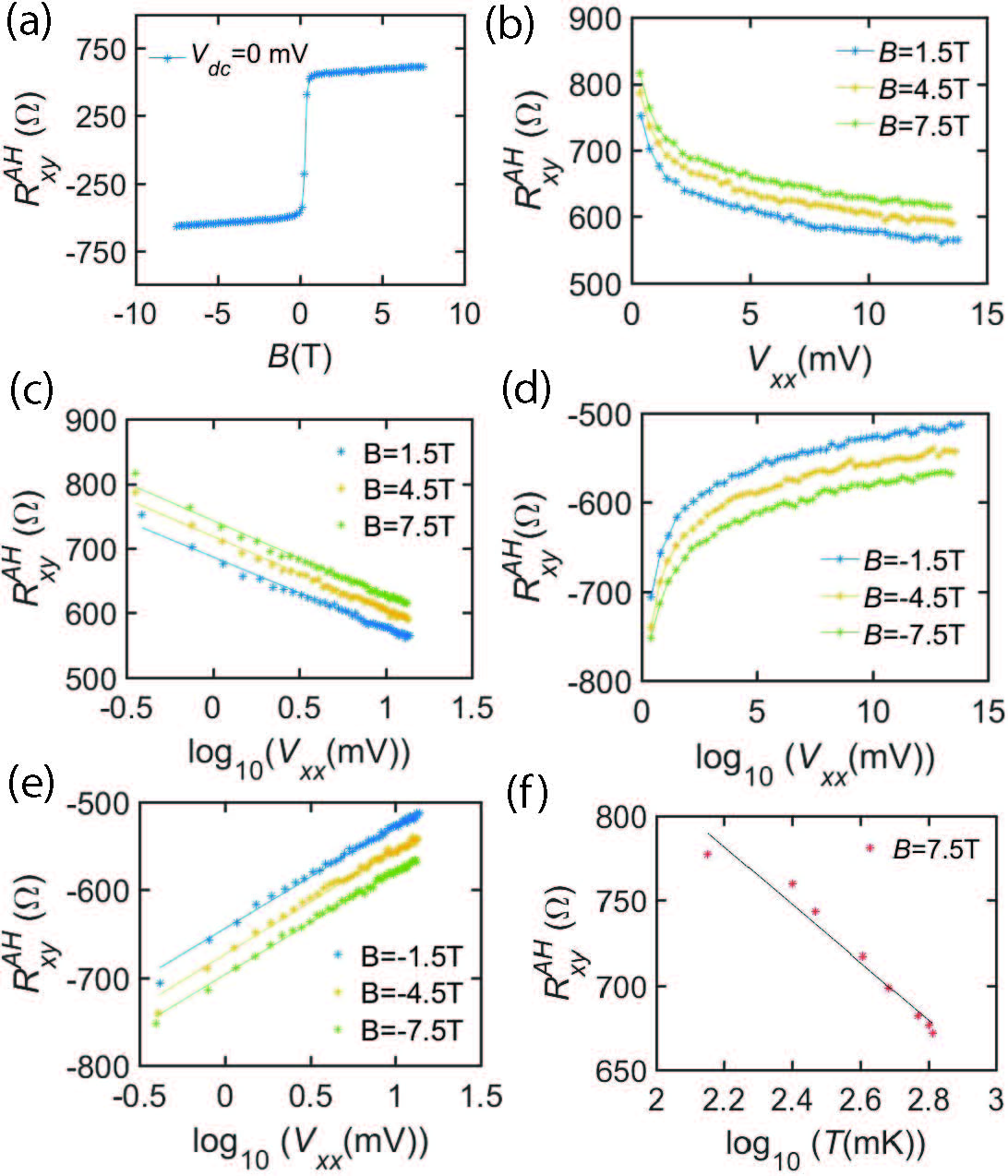}
\end{centering}
\caption{(color online) Voltage bias and temperature dependence of anomalous Hall effect of V-doped 4 QL thick (Bi,Sb)$_2$Te$_3$ films. (a) Magnetic field dependence of anomalous hall resistance $R^{AH}_{xy}$ measured at $V_{dc} = 0$. (b) Bias dependence of  $R^{AH}_{xy}$ at $B$ = 1.5, 4.5 and 7.5 T. (c) Logarithmic plot for bias dependence of  $R^{AH}_{xy}$   at $B$ = 1.5, 4.5 and 7.5 T.  (d) Bias dependence of  $R^{AH}_{xy}$ at $B$ = -1.5,- 4.5 and -7.5 T. (e) Logarithmic plot for bias dependence of  $R^{AH}_{xy}$  at $B$ = -1.5, - 4.5 and -7.5 T. (f) $R^{AH}_{xy}$ exhibits a logarithmic dependence on temperature at  $B$ = 7.5 T in the temperature range of 140 mK to 650 mK .}
\label{transverse resistance}
\end{figure}

Weak localization in disordered two dimensional (2D) systems is also a potential explanation for logarithmic increase in resistance at low temperatures. The existence of weak localization relies on the existence of coherent constructive interference of time reversed trajectories for an electron to return to the origin~\cite{Bergmann1984}. Moderate external magnetic fields as well as magnetic impurities, that break time reversal symmetry, are typically enough to suppress logarithmic corrections arising from weak localization~\cite{Altshuler80,RMP85,AABook}. However, the logarithmic corrections observed in our experiments persist even at fields of 7.5 T.

A way to identify logarithmic corrections due to weak localization is by the absence of logarithmic corrections to $R_{xy}$  ~\cite{Fukuyama1980,Altshuler80,RMP85}. In  Fig. \ref{transverse resistance}(a) anomalous Hall measurements are shown without an applied bias. The $R^{AH}_{xy}$ jumps at the coercive field when the magnetization switches its direction. Fig. \ref{transverse resistance}(b)- Fig. \ref{transverse resistance}(c) shows that the logarithmic dependence on voltage bias are present in anomalous Hall resistance  $R^{AH}_{xy}$  as well. The data is antisymmetric in magnetic field as shown in  Fig. \ref{transverse resistance}(d)- Fig. \ref{transverse resistance}(e) because of which we can rule out spurious  $R_{xx}$  contributions. Interestingly, if instead of bias voltage temperature of the sample is changed, similar decrease in resistance $R^{AH}_{xy}$  is observed as shown in Fig. \ref{transverse resistance}(f). Therefore, weak localization cannot explain the transport behavior that we observe.

Logarithmic corrections to conductance could also arise from scattering off magnetic impurities as in the Kondo effect~\cite{Anderson1961, Kondo1964, Appelbaum1966}.  However, in the ferromagnetic state  magnetic spin-flips should become increasingly energetically unfavorable at low temperatures and at large external magnetic fields. More importantly, the fact that logarithmic dependences are also observed in topological insulator thin films  in the absence of magnetic dopants \cite{Liu2011,Wang2011} makes this scenario an unlikely explanation of our findings.


Magneto-transport studies  in pristine topological insulator Bi$_2$Se$_3$ films found it crucial to include electron-electron interactions \cite{Liu2011,Wang2011}.
We explain why the observed logarithmic singularities are due to electron-electron interactions in the 2D surface states. As first realized by Altshuler and Aronov \cite{AA1979} (AA) disordered 2D electron systems exhibit a breakdown of the Fermi-Liquid theory due to reduced ability of the disordered electron gas to screen the Coulomb interaction. The logarithmic corrections in the AA theory are pervasive and are expected to arise not only in transport properties but also in equilibrium thermodynamic quantities such as specific heat~\cite{RMP85}. One of the key differences of the AA corrections with those in the localization theory is that both the longitudinal and Hall resistivities are expected to acquire logarithmic corrections \cite{RMP85,Altshuler80}. In fact the logarithmic corrections are most easily expressed in terms of conductivities rather than resistivities in the AA theory, because the Hall conductivity is expected to remain unchanged. Specifically one expects the logarithmic corrections in the AA model to be given by \cite{Altshuler80,RMP85,AABook}:

\be\label{sigma}
\delta\sigma_{xx}(\varepsilon)=\kappa  \frac{e^2}{h} \log (\frac{\varepsilon \tau}{\hbar}), \ \delta \sigma_{xy}=0,
\ee

\noindent where $\tau$ is the elastic scattering time, and $\varepsilon$ is an appropriate energy scale that can be chosen to be the largest among the temperature $k_B T$ or the frequency $\hbar \omega$, at which the conductivity is probed. $\kappa$ is a dimensionless number that takes different values for spinless and spinful electrons, and depends on a dimensionless parameter $F$ that characterizes a Hartree contribution to the conductivity corrections~\cite{altshuler1982effects,RMP85,AABook}. This parameter takes the following forms for spinless and spinful electrons:

\be\label{Fless}
\kappa^{\rm spinless}=\frac{1}{2\pi},
\ee

\be\label{Ffull}
\kappa^{\rm spinfull}=\frac{1}{2\pi} (2-2F),
\ee

\noindent For short range interaction, $F$=1 and long range interaction $F$=0 \cite{altshuler1980interaction}.  For spin-split bands one expects that for a spin splitting $\Delta\gg  k_BT $, the only singular logarithmic terms arise from exchange and $S_z=0$ Hartree contributions, and the expression for $\kappa$ is \cite{RMP85}:

\be\label{Fpart}
\kappa^{\rm spin-split}=\frac{1}{2 \pi} \left(2-F\right)  \approx  0.32 \left(1-\frac{1}{2} F \right).
\ee

\noindent Our magnetic samples are expected to be spin-split, whereas the precise level of spin polarization is unknown to us ~\cite{RMP85,AABook}.

\begin{figure}[h!]
\begin{centering}
\includegraphics[width=8.5cm]{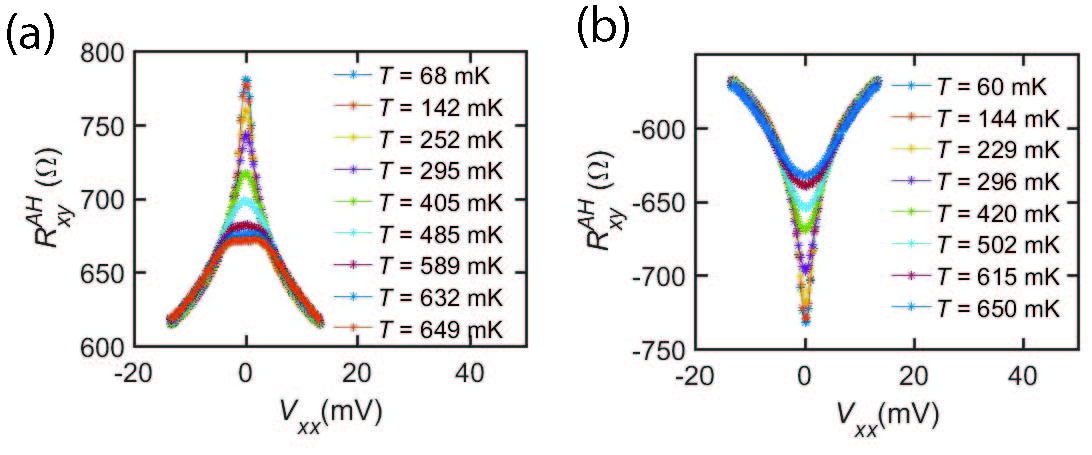}
\end{centering}
\caption{(color online) Voltage bias dependence of the anomalous Hall effect  at several fixed temperatures in V-doped 4 QL thick (Bi,Sb)$_2$Te$_3$ films for (a) $B$= 7.5T and (b) $B$= -7.5T. }
\label{AHE}
\end{figure}

Fig. \ref{AHE}(a)-(b) further explores the dependence of anomalous Hall effect on both voltage bias and temperature. Increasing either voltage or temperature lowers the anomalous Hall resistance which supports a self-heating mechanism due to applied bias.

\begin{figure}[h!]
\begin{centering}
\includegraphics[width=8.5cm]{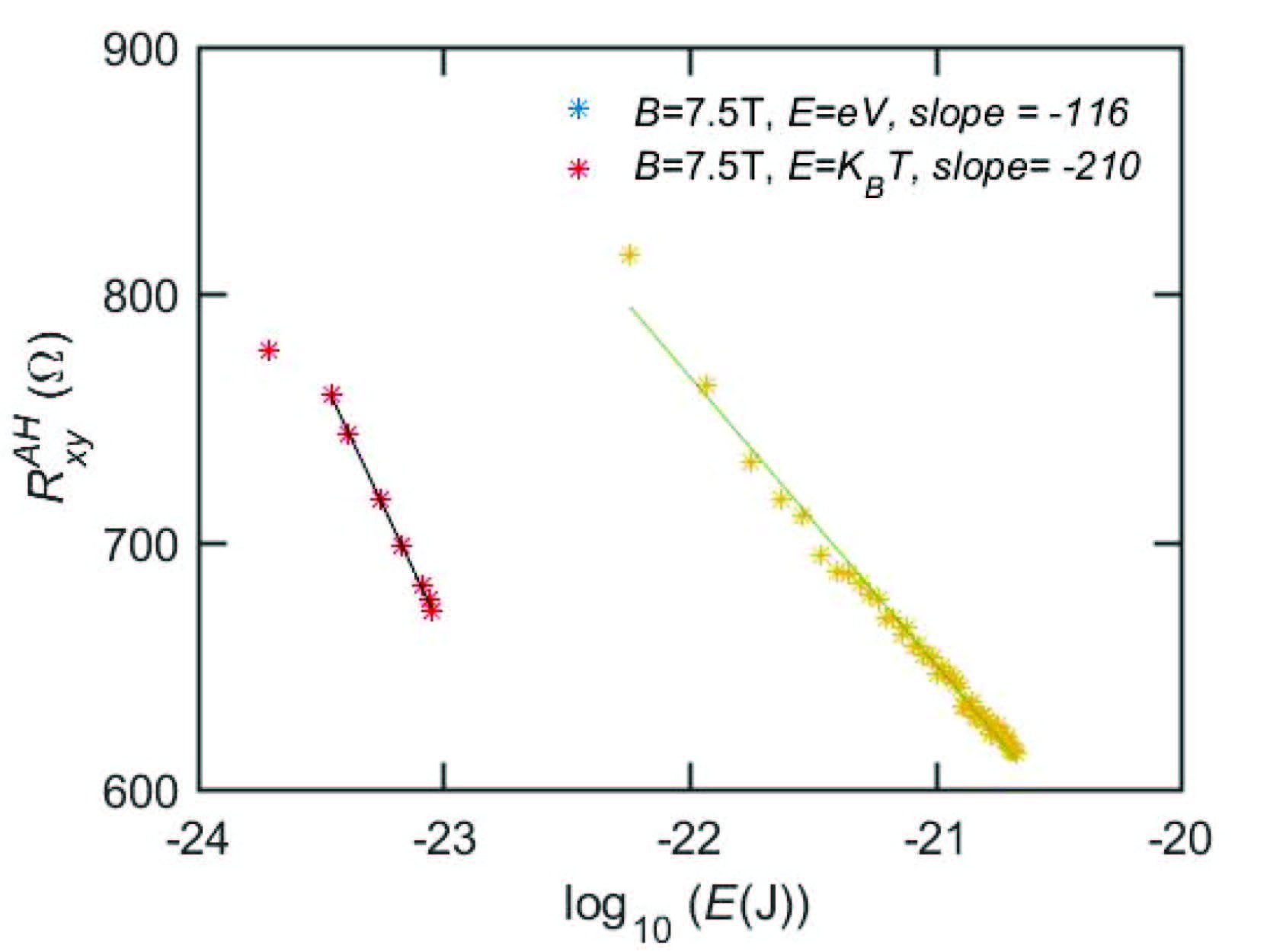}
\end{centering}
\caption{(color online) Comparison of the dependence of  anomalous Hall resistance $R^{AH}_{xy}$ on logarithm of $K_BT$ and $eV$ in V-doped 4 QL thick (Bi,Sb)$_2$Te$_3$ films at $B$= 7.5 T. }
\label{power}
\end{figure}

We believe that the origin of the non-Ohmic behavior we observe, namely the logarithmic dependence of the conductivity on voltage bias, is fundamentally no different than the logarithmic dependence on temperature and can be understood simply as a consequence of Joule heating. In other words, as the electrons are accelerated by the electric field they inevitably gain energy, and, once they reach a steady state of current flow, this inevitably implies that the electrons posses a larger effective temperature compared to that of the lattice or other reservoirs that serve as heat sinks. By appealing to a simple model of Joule heating \cite{Joule} one can effectively replace the argument of the logarithm in Eq. (1) by $\varepsilon \sim$ max$(AV^{2/(2+p)},K_BT, \hbar\omega)$, where $V$ is the voltage bias that drives the transport, $A$ is a constant, and $p$ is the power that controls the temperature dependence of the electron's inelastic scattering rate, $\tau_{in} \propto T^{-p}$ . The logarithmic fits of the Hall resistivity vs temperature have approximately twice the slope of those Hall resistivity vs the bias voltage indicating that $p\sim 2$, as shown in Fig. \ref{power} .

The Joule heating induced by the bias voltage results to be a more efficient way to tune the electron temperature than the direct control of the temperature of the sample, and, hence we will focus on this dependence for the remainder of the discussion. The expected behavior of the correction to the conductivity for low temperatures dc measurements from the AA theory as a function of voltage is:

\be\label{sigmaV}
\delta\sigma_{xx}(V) = \frac{2\kappa}{(2+p)}\frac{e^2}{h} \log (V), \ \delta\sigma_{xy}(V)=0.
\ee

\begin{figure}[h!]
\begin{centering}
\includegraphics[width=8.5cm]{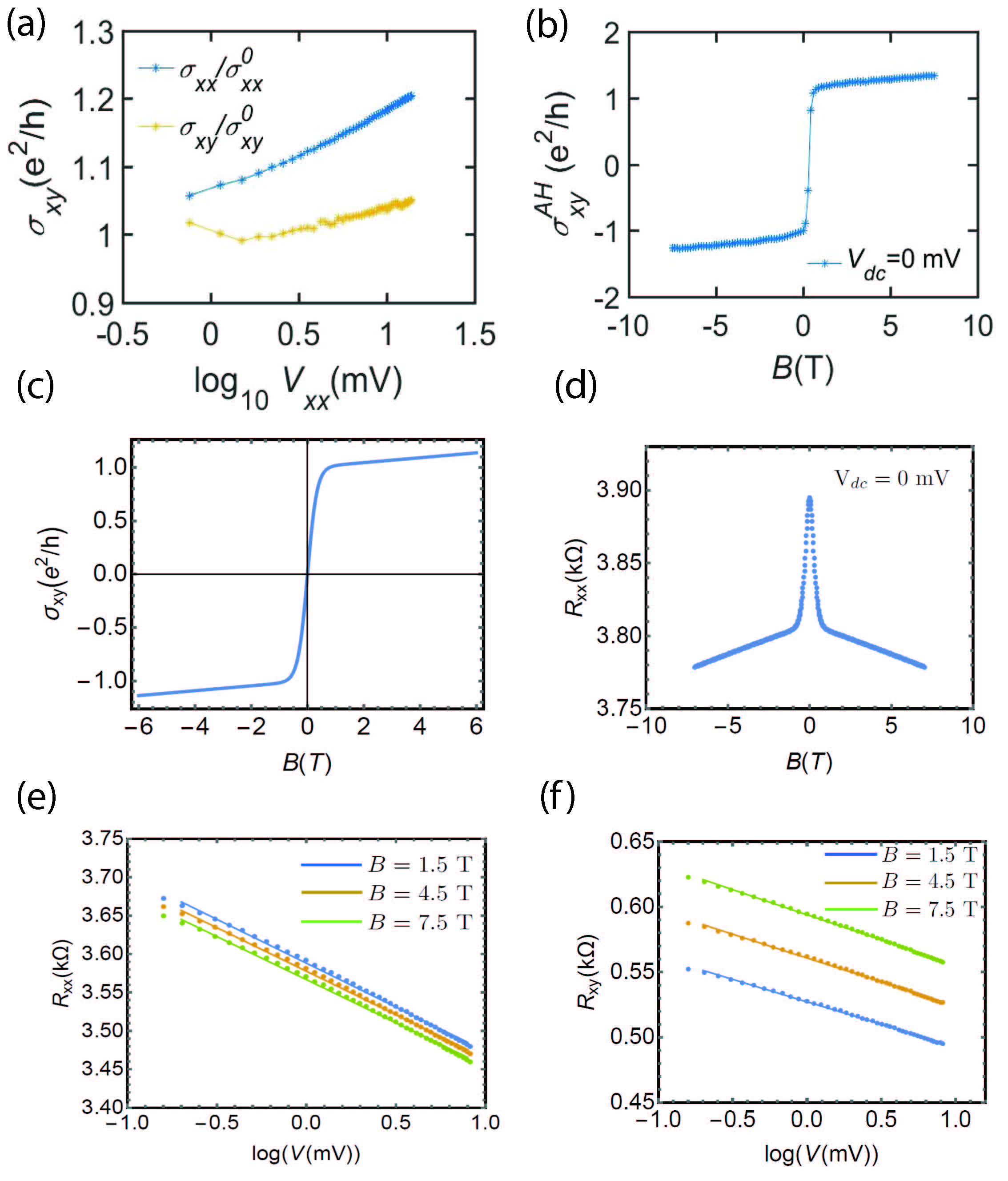}
\end{centering}
\caption{(color online) Transport coefficients in the V-doped Hall bar (device H$_1$) as a function of bias. (a) Logarithmic plot for $\sigma_{xx}$ (blue) and $\sigma_{xy}$ (orange) at $B=$ 1 T. (b) Magnetic field dependence of $\sigma_{xy}$ at $V_{dc}$=0. (c),(d) Theoretical model of the magnetic field dependence of $\sigma_{xy}$ and $R_{xx}$.  (e),(f) Simulation of dependence of $R_{xx}$ and $R_{xy}$ on bias at $B$=1.5 T, 4.5 T and 7.5 T as expected from interaction corrections in disordered 2D films.}
\label{RvsSigma}
\end{figure}

From the data, $\sigma_{xx}$ and $\sigma_{xy}$ is calculated using the relations

\be\label{sigmaV}
\sigma_{xx}=\frac{\rho_{xx}}{\rho_{xy}^2 +\rho_{xx}^2},
\sigma_{xy}=\frac{\rho_{yx}}{\rho_{xy}^2 +\rho_{xx}^2}
\ee

\noindent where $\rho_{xx}$ and $\rho_{xy}$ are the resistivities considering the square shaped sample geometry. As illustrated in Fig. \ref{RvsSigma}(a), our data is consistent with logarithmic corrections in $\sigma_{xx}$ while no apparent logarithmic corrections in $\sigma_{xy}$, as expected from the AA theory. The anomalous Hall conductivity is nearly quantized at +/-  $e^2/h$ as shown in Fig.  \ref{RvsSigma}(b). Anecdotally, we argue that quantization of $\sigma^{AH}_{xy}$ is less sensitive to finite bulk carriers than $R^{AH}_{xy}$.

To quantify the logarithmic behavior we fit the voltage-dependent nonlinear conductivity as (expressing bias in volts):

\be
\sigma_J(V) = \sigma^0_J+ \delta \sigma_J \log(V), \ J=\{xx,xy\}
\ee

\noindent we obtain $\sigma^0_{xx} \approx 9.17 e^2/h$,  $\delta \sigma_{xx} \approx 0.33 e^2/h$, and $\sigma^0_{xy} \approx 1.58 e^2/h$,  $\delta \sigma_{xy} \approx 0.02 e^2/h$. Notice the smallness of the bias dependence of $\sigma_{xy}$ compared to $\sigma_{xx}$. Therefore, considering that it is possible that small systematic errors can arise from mixing of $R_{xx}$ and $R_{xy}$ (e.g. if contacts are sligthly misaligned $R_{xy}$ picks a small contribution from $R_{xx}$), we conclude that our data is consistent with $\sigma_{xy}$ with negligible logarithmic bias dependence and while having significant logarithmic bias dependence on $\sigma_{xx}$, as expected from the AA theory.  We observe, however, an interesting quantitative deviation from the expectation of the AA theory. Using the approximate value of $p\sim2$, obtained by comparing the temperature and the voltage fits (see Fig. \ref{power}), the fitted parameter $\kappa$ reads as:

\be
\pi \kappa_{\rm fit} \sim 2.
\ee

\noindent However, from Eqs.\eqref{Fless}-\eqref{Fpart}, we expect $\pi \kappa \leq 1$, under the natural assumption of repulsive interactions $F>0$. The origin of this discrepancy is at present unknown to us, but we wish to remind the reader that the equations of the AA we have employed were derived for parabolic electrons without Berry phase effects, and, it remains to be determined whether nontrivial orbital coherence, such as those giving rise to Berry curvatures for the bands of interest here, affect in any way the classic results of the AA theory.


A simple modeling of the resistivity can be done by using the expected conductivity behavior from the AA theory. The resistivity is taken to be of the form: $\sigma_{xx}=\sigma_{xx}^0+\delta \sigma_{xx}^0 \log(|V|+V_0)$, where $V_0\sim 0.4$ mV is essentially a cutoff of the logarithm at small bias (which is controlled by the temperature scale $T_0$ and the constant $A$ in the Joule heating model), and $\sigma_{xx}^0$ and $\delta \sigma_{xx}^0$ are field and bias independent quantities obtained by linear fitting of the logarithmic plots of the conductivity \cite {suppl}. We add a simple description of the anomalous hall effect in which the Hall conductivity has a jump of $e^2/h$ near zero applied magnetic field in addition to the usual linear term reflecting the classical Hall effect. $\sigma_{xy}$ in the model is presented in  Fig. \ref{RvsSigma}(c) and has the form: $\sigma_{xy}=\frac{e^2}{h}\tanh(B/B_0)+\delta \sigma_{xy}^0 B$, where $B_0\sim 0.3$ T reflects broadening of the jump of the magnetization as a function of field and $\delta \sigma_{xy}^0$ is field and bias independent. The model is able to reproduce the essential behavior of the resistivities and it is shown in Fig. \ref{RvsSigma}(d) - Fig. \ref{RvsSigma}(f).

\begin{figure}[h!]
\begin{centering}
\includegraphics[width=8.5cm]{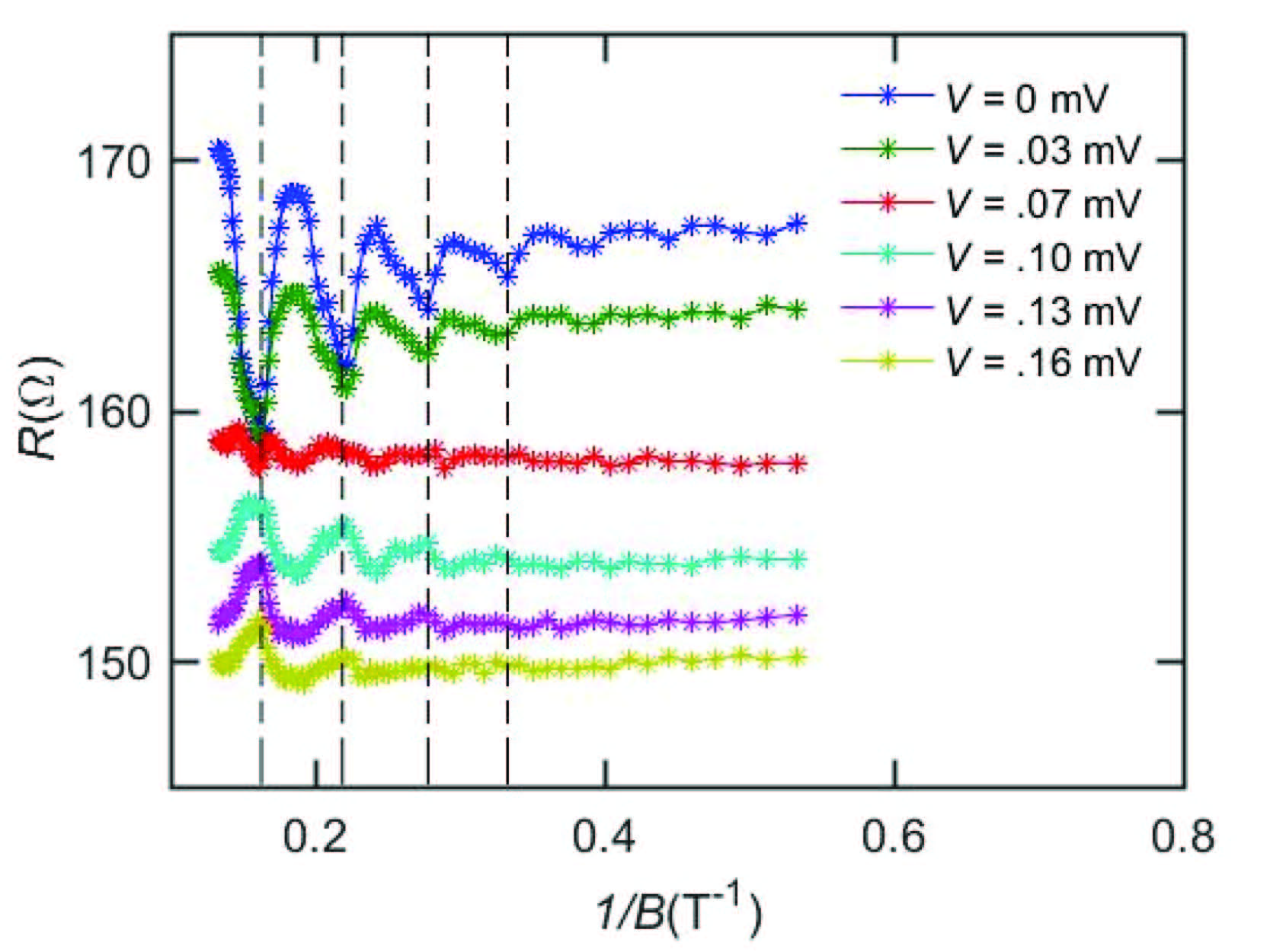}
\end{centering}
\caption{(color online) Voltage bias dependence of quantum oscillations from magnetoresistance in V-doped (Bi,Sb)$_2$Te$_3$ device D$_1$ of width $W$=0.2 $\mu$m.}
\label{Shubnikov-de Haas}
\end{figure}

The  transport results discussed above are for larger Hall bar ($\sim$ 20 $\mu$m) samples. Interestingly, when the device dimension was reduced to submicron range, prominent Shubnikov-de Haas (SdH) oscillations were observed. For example, in a 0.2 $\mu$m wide device (device D$_1$) measured by two terminal the oscillations  were periodic in 1/B and have a non-trivial dependence on bias voltage. These quantum oscillations were seen in multiple samples with Ti/Nb/NbN and Ti/Al contacts. In particular, the zero bias minima turn into maxima in resistance at large voltage bias as shown in Fig. \ref{Shubnikov-de Haas}.

\begin{figure}[h]
\begin{centering}
\includegraphics[width=3.3in]{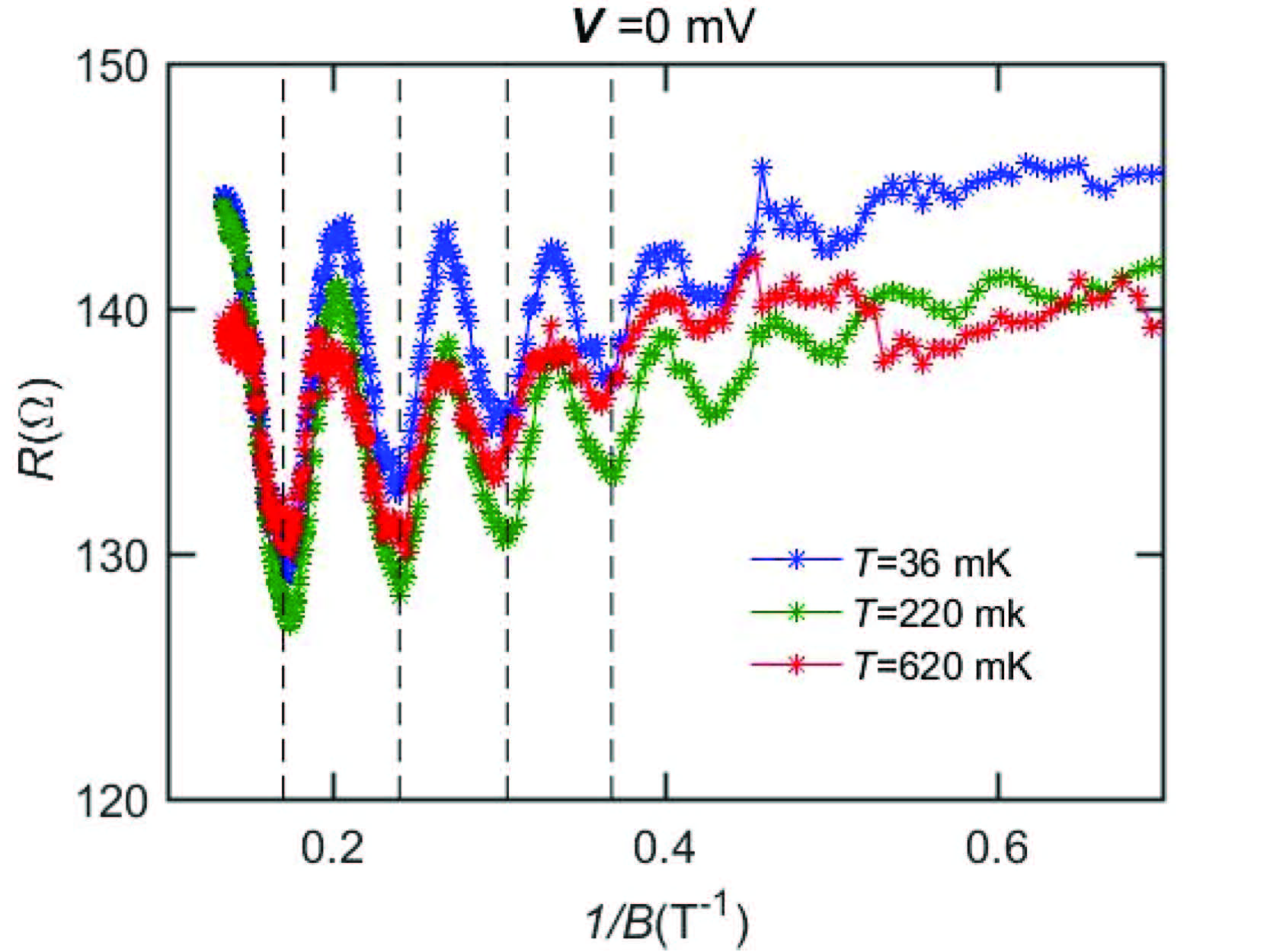}
\end{centering}
\caption{(color online) Temperature dependence of quantum oscillations from magnetoresistance in V-doped (Bi,Sb)$_2$Te$_3$ device D$_1$ of width $W$=0.2 $\mu$m.}
\label{Temperature_dependence}
\end{figure}

We also studied the effect of temperature on the magneto-oscillations as shown in Fig. \ref{Temperature_dependence}.  The amplitude of the quantum oscillations is found to decrease with increase in temperature. However, the transition from maxima to minima could not be observed at temperatures accessible in the dilution fridge. The inferred electron density is 9 $\times$ 10$^{11}$ cm$^{-2}$ (4.5 $\times$ 10$^{11}$ cm$^{-2}$) for spinful (spinless) Fermions. The period of the SdH could not be changed by applying a backgate or topgate. The screening of the top and bottom gates by the surface states results in inability to change the Fermi energy of the bulk states as has been observed in other topological materials \cite{Mahler2017}. An estimate of the electron gas mobility is made from the onset magnetic field of the SdH oscillations $\mu_q\approx\frac{1}{B_q}\approx 6,000$ cm$^2$ V$^{-1}$s$^{-1}$ for the 200 nm wide device \cite{cui2015multi}. This mobility is intriguingly large compared to macroscopic samples.

\begin{figure}[h!]
\begin{centering}
\includegraphics[width=8.5cm]{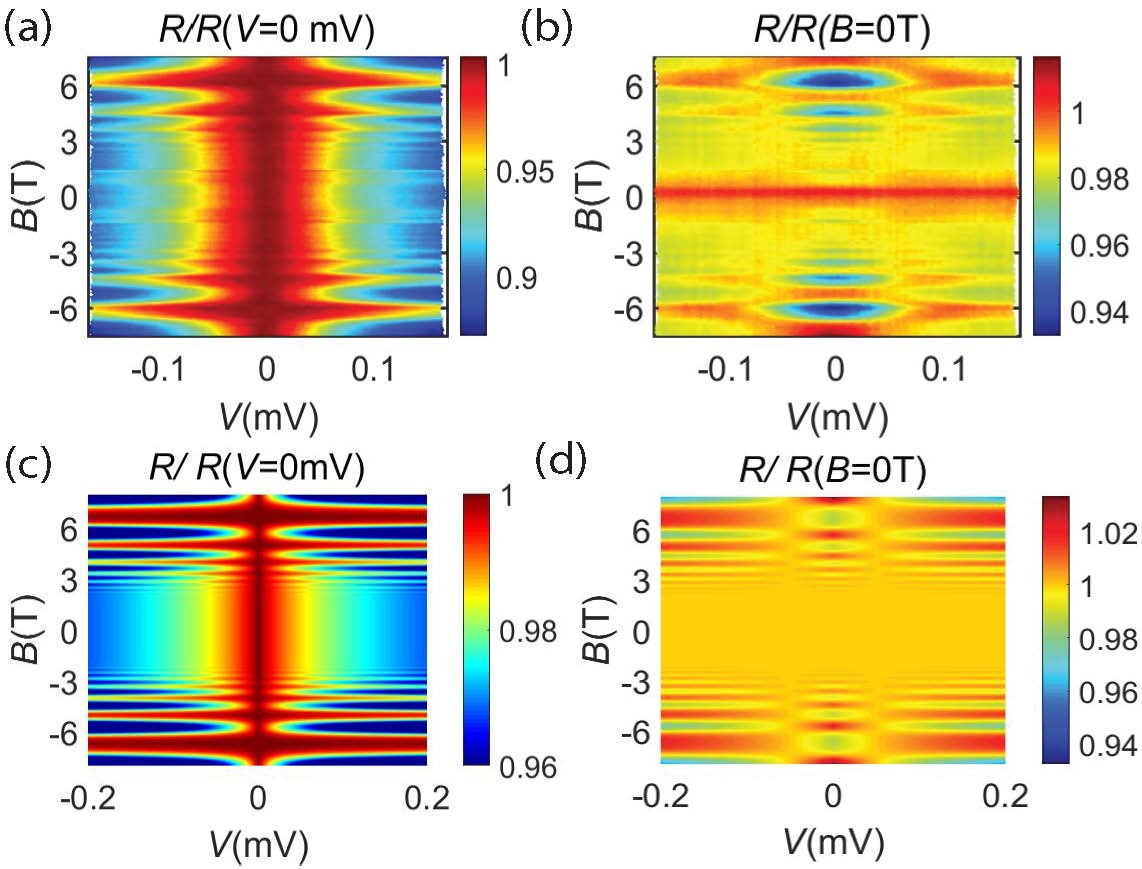}
\end{centering}
\caption{(color online) (a),(b) The resistance in V-doped (Bi,Sb)$_2$Te$_3$ device D$_1$ is normalized to its zero bias value and normalized to  its zero magnetic field value respectively. The evolution is studied with magnetic field and applied bias voltage. The SdH oscillations are present both at small and large bias voltages. However, the zero bias maxima  become minima at large bias voltages at a fixed magnetic field and vice versa. (c) A  model with two conduction mechanisms in parallel that incorporates logarithmic decay with applied bias of the SdH oscillations on top of a constant  background  conduction. The resistance is normalized to the zero bias value for comparison to the experimental data. (d) The two conduction model captures the evolution from maxima to minima of the resistance normalized to the zero magnetic field value with voltage bias.}
\label{Shubnikov-de Haas Model}
\end{figure}

We describe a transport model for the non-trivial dependence of the quantum  oscillations on voltage bias that we have observed in the narrow junctions. We assume the magneto-conductance to be given by
\begin{equation}
g(B,V)=g_0+ \alpha \rho(\epsilon_F,B)\ln({\varepsilon\tau/\hbar})
\end{equation}
where  $g_0$ is assumed to be a constant background conduction and $\rho(\epsilon_F,B)$ is the SdH density of states given by
\begin{equation}
\rho(\epsilon_F,B)=\frac{2eB}{h}\sum_{j=0}^{j=\infty}\frac{1}{(2\pi)^{1/2}\Gamma} \exp({-\frac{(\epsilon_F-\epsilon_j)^2}{2\Gamma^2}}){}
\end{equation}
where $\Gamma$ is the half-width of  Landau level broadening and $\epsilon_j$ is the single particle Landau level energy \cite{Eisenstein1985}.
Presence of both bulk and surface conduction mechanisms in topological insulators has been found previously \cite{Analytis2010,James2010,Qu2010,Xiong2012,Cao2012,Saha2014}. Fig. \ref{Shubnikov-de Haas Model}(a) and Fig. \ref{Shubnikov-de Haas Model}(b) display the magneto-transport data normalized to zero bias and zero magnetic field. Similarly, in the model, the resistance $ r(B,V) = 1/g(B,V)$ is normalized to its value at zero bias voltage and magnetic field as is shown in Fig. \ref{Shubnikov-de Haas Model}(c) and Fig. \ref{Shubnikov-de Haas Model}(d) respectively.

While we do have an understanding of the voltage dependence of the oscillations, there are properties  that are less well understood.  The contrast of the quantum oscillations is found to decrease systematically with increasing width. Such dependence of visibility of quantum oscillations on channel width is unusual. The quantum oscillations are discussed in further detail in supplementary materials \cite {suppl}.

\section{CONCLUSION}

We have studied magnetotransport in V-doped (Bi,Sb)$_2$Te$_3$ and find logarithmic singularities in longitudinal resistance $R_{xx}$ and anomalous Hall resistance  $R^{AH}_{xy}$ which is well explained quantitatively by quantum corrections due to electron-electron interactions. In submicron scale devices, SdH oscillations are observed where the maxima transition to minima with voltage bias. A simple transport model explains these observations.

\section{ACKNOWLEDGEMENTS}
The authors acknowledge insightful discussions with Ashwin Viswanathan, Liang Fu, Brian Skinner, Toeno Vander Saar, Lucas Orona and Anindya Das. The authors are grateful to Di S. Wei, Tony X. Zhou, Pat Gunman and Andrei Levin for invaluable help with developing the sample fabrication. AY, DN, KS and JW acknowledge the support from Gordon and Betty Moore Foundation Grant No. 4531, ARO Grant No. W 911 NF-16-1-0491, DOE Grant No. DE-SC0001819 and ARO Grant No. W911NF-17-1-0023. AY also acknowledges support in part by the U. S. Army Research Laboratory and the U. S. Army Research Office under Grant No. W911NF-16-1-0491.
 IS acknowledges support from MIT Pappalardo Fellowship. PK, GHL and KH acknowledge support from NSF Grant No. DMR-1420634. JSM, CZC and YO acknowledge the support from NSF Grant No. DMR-1700137, ONR Grant No. N00014-16-1-2657.

\section{Supplemental Materials for 'Logarithmic singularities and quantum oscillations in magnetically doped topological insulators'}
\subsection{Processing of (Bi,Sb)$_2$Te$_3$ on SrTiO$_3$ substrate}

\begin{figure}[h]
\begin{centering}
\includegraphics[width=3.3in]{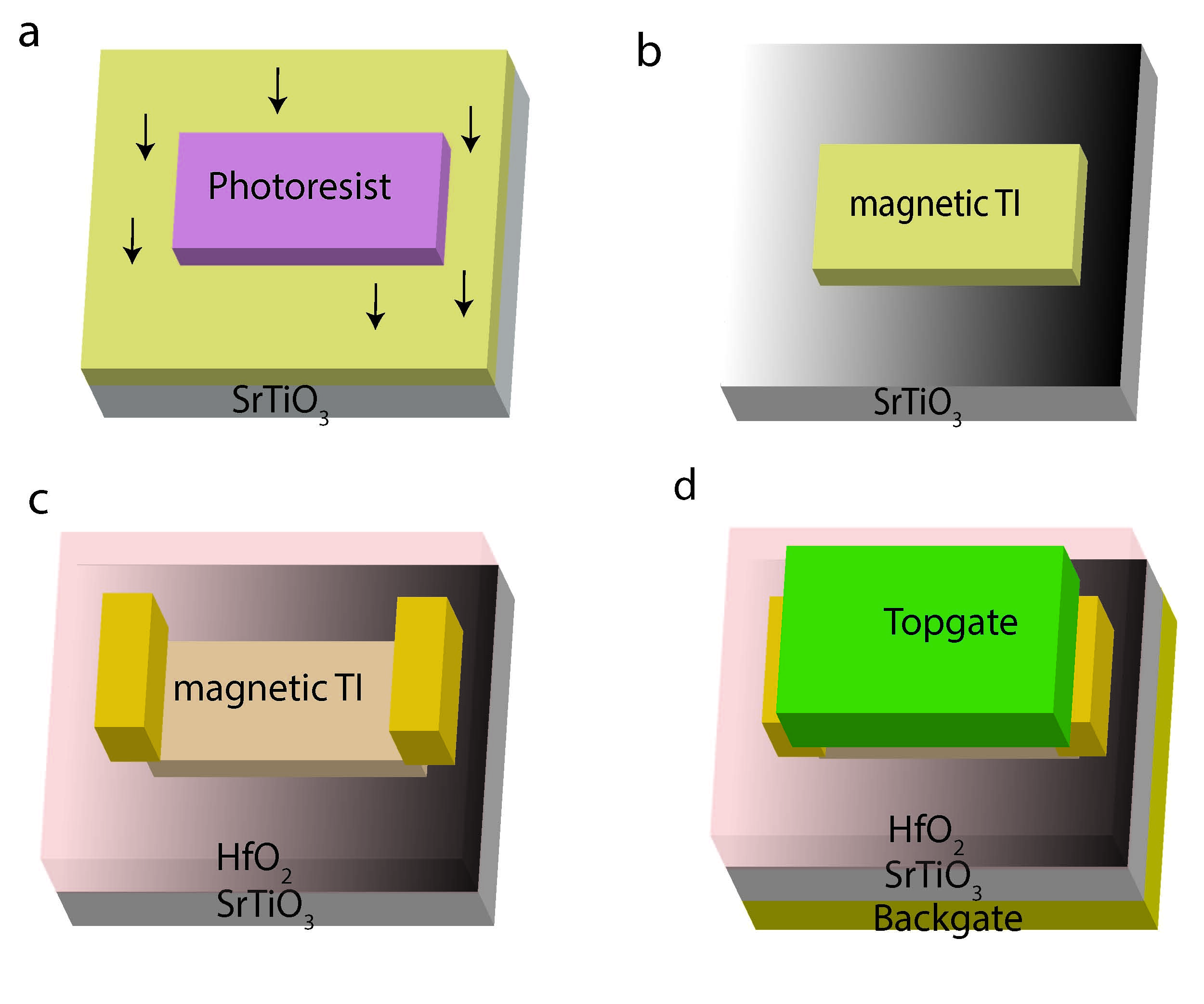}
\end{centering}
\caption{Schematic of the ebeam fabrication process of topological insulators. (a) The etch mask is patterned photolithographically from photoresist S1813 by UV exposure. (b) The unprotected topological insulator film is removed by oxygen plasma. (c) Superconducting or normal metal contacts are ebeam lithographically patterned and metallized. Subsequently, a high $\kappa$ dielectric of HfO$_2$ is grown by atomic layer deposition. (d) A topgate as well as a backgate of Ti/Au is deposited.}
\label{Processing}
\end{figure}

We developed ebeam fabricated devices of pristine and magnetic topological insulators. This included several challenges including SrTiO$_3$ being highly insulating, (Bi,Sb)$_2$Te$_3$ degrading at elevated temperatures and oxidation of the sample without proper capping layer. Here we report the details of sample processing which would allow several design exploration in the field of topological insulators. The device processing illustrated in Fig.  \ref{Processing} compares favourably to scratched Hall bar devices and does not effect the film quality.

\subsubsection{Recipe for etching }
To etch 4 quintuple layers of pristine and V-doped (Bi,Sb)$_2$Te$_3$  with 10 nm Te capping layer, an etch mask of photoresist S1813 is used. Using a laser writer, the mask pattern is defined followed by development in developer CD-26. A 5 mins O$_2$ plasma etch removes the topological insulator film exposing the transparent SrTiO$_3$.
\subsubsection{ Making electrical contact} SrTiO$_3$ is insulating because of which charging effects are a major concern for high resolution lithography. For patterning the contacts by ebeam lithography, the resist layers of PMMA C6 495 K and PMMA A4 950 K are baked at 100 $^{\circ}$ C as higher temperatures lead to sample degradation. Then 10 nm of Al is deposited for reducing charging effects which is transparent under scanning electron microscope imaging.  After ebeam exposure, the Al is removed in AZ 400K developer in 5 mins and then developed in MIBK : IPA (1 : 3). As the topological insulator film is protected from oxidation by an insulating capping layer, an in-situ argon plasma etch is done before deposition of Ti(10 nm)/Au(100 nm) or Ti (10 nm)/Nb(5 nm)/NbN(50 nm). Optimization of the etch time is necessary to make good electrical contact to the film.
\subsubsection{ Making topgated devices} After electrical contacts are made, a 20 nm HfO$_2$ dielectric is grown by atomic layer deposition (ALD) at 80 $^{\circ}$ C.  A Ti (10 nm)/Au(100 nm) topgate is deposited on top of the HfO$_2$ dielectric. The topgated devices can be effectively tuned through the charge neutrality point.
\subsubsection{ Making backgates} After the topside fabrication is complete, the device is coated with PMMA C6 495 K baked at 100  $^{\circ}$ C and flipped over to deposit Ti(10 nm)/Au(100 nm) backgate. The protecting ebeam resist is then dissolved and the device is wirebonded.

\subsection{Measuring the contact resistance by transfer length method}

The two terminal resistance to the pristine and magnetic topological insulator film is measured as a function of various gap sizes $W$. By extrapolating to gap size $W$ = 0, the residual resistance is measured. The equation used to determine the contact resistances by transfer length method are \cite{Venugopal2010,Wang2013}:
\begin{equation}
R_{total}=\rho_{sheet}\frac{W}{d} + 2R_c
\end{equation}
where $\rho_{sheet}$ is the sheet resistance. In this technique, we conclude that the contact resistance is below 50 $\Omega$ at 4.2 K and makes good electrical contact to the film.

\subsection{Topgating of magneto-oscillations}

\begin{figure}[h]
\begin{centering}
\includegraphics[width=3.3in]{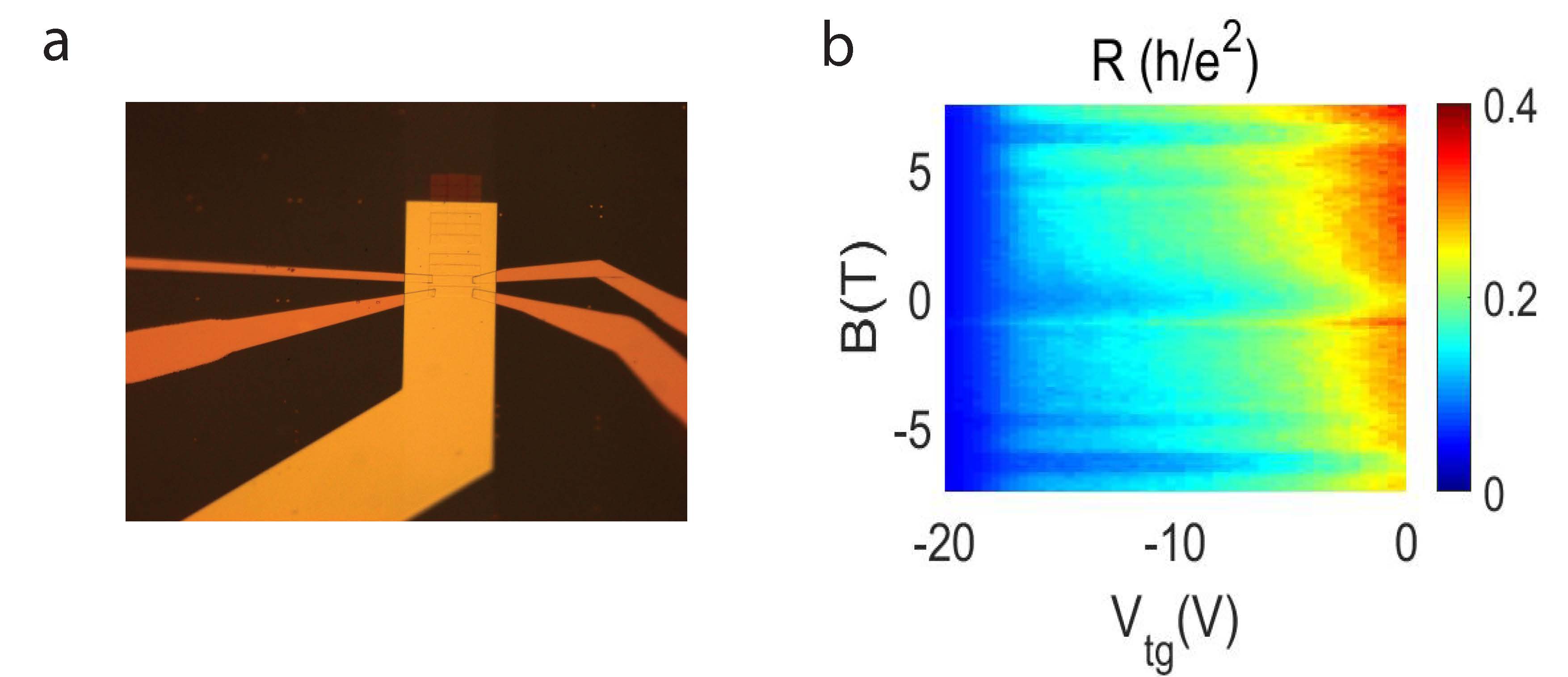}
\end{centering}
\caption{(a) Optical microscope image of topgated topological insulator film with 20 nm HfO$_2$ dielectric. (b) Two terminal resistance in Ti/Nb/NbN devices show significant topgate response. Intriguingly, the magneto-oscillations in $B$-field do not depend on changes in topgate voltages.}
\label{Gate_response}
\end{figure}

We integrate ALD grown high-k dielectric on pristine and magnetic topological insulators which allows to tune the Fermi-energy. The ALD deposition of 20 nm  HfO$_2$ dielectric  is done at 80$^{\circ}$ C to avoid any sample degradation due to overheating. All topgated devices could be tuned through charge neutrality. Intriguingly, the magnetic field dependence of the quantum oscillations was independent of the topgate voltage as shown in  Fig. \ref{Gate_response}.

\subsection{Estimating density and mean free path}

\begin{figure}[h]
	\begin{centering}
		\includegraphics[width=3.3in]{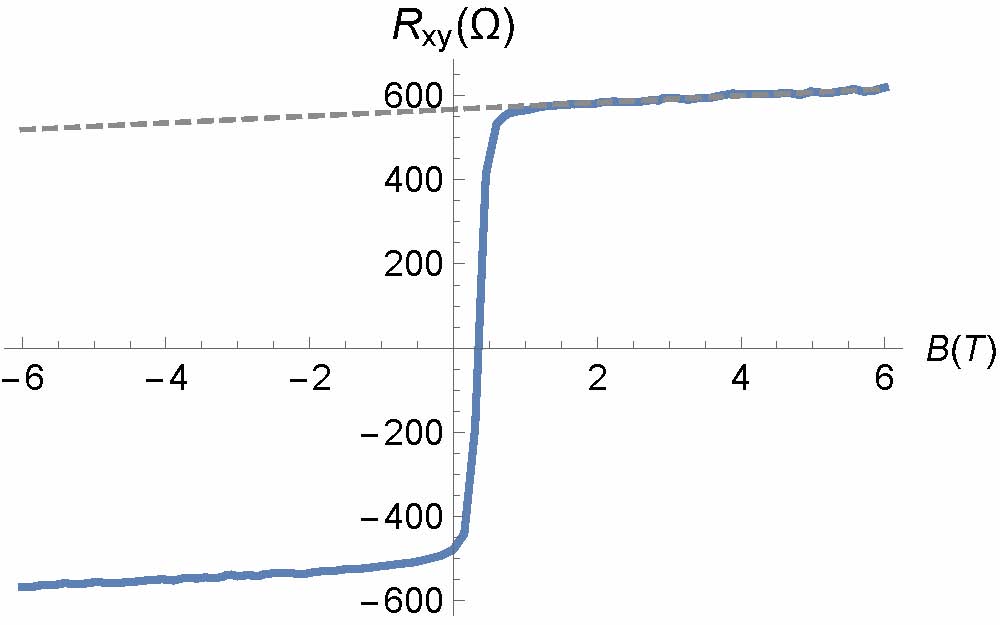}
	\end{centering}
	\caption{(color online) $R_{xy}$ for Hall-bar sample as a function of magnetic field.}
	\label{Rxy1}
\end{figure}

\noindent Fig.~\ref{Rxy1} shows the Hall resistance $R_{xy}$ for Hall-bar sample (taken near zero bias) as a function of magnetic field. By making a linear fit to the large field values (shown as a dashed line) we can estimate the sample density using the semiclassical value of the Hall resistivity: $R_{xy}= B/nec$. From this we infer a carrier density of $n \approx 8.3 \times 10^{17} m^{-2}$.

\begin{figure}[h]
	\begin{centering}
		\includegraphics[width=3.3in]{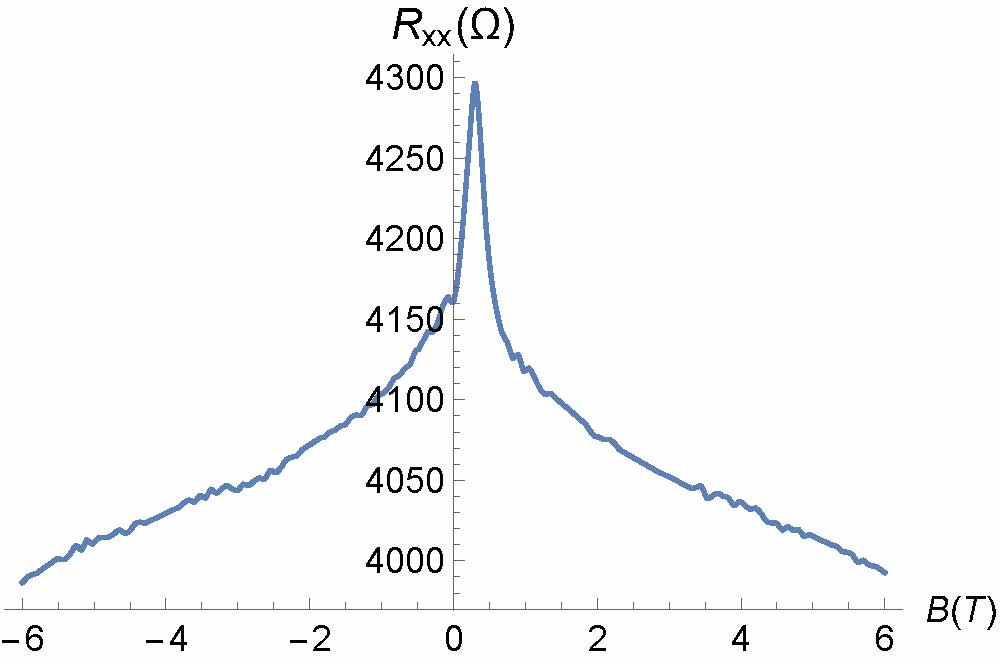}
	\end{centering}
	\caption{(color online) $R_{xx}$ for Hall-bar sample as a function of magnetic field.}
	\label{Rxx1}
\end{figure}

\noindent To estimate the elastic mean free path we use the values of the longitudinal resistance $R_{xx}$ for the Hall-bar sample (taken near zero bias) as depicted in Fig.~\ref{Rxx1}. using the formula:

\be
k_F l=\frac{2 R_K}{g \rho_{xx}},
\ee

\noindent where $R_K$ the Von-Klitzing's constant $ R_K=25k \Omega$ and $g=\{1,2\}$ for spinless/spinfull electrons. Assuming the sample to be squared (we only want a rough order of magnitude for the mean free path), and taking $R_{xx} \sim 4.3 k\Omega$, we obtain $k_F l \sim \{12,6\}$ for spinless/spinfull electrons. So, we expect an elastic mean free path of about $l\sim \{3.7nm,2.6nm\}$. This estimate implies that the sample is rather clean, and perturbative calculations on $1/(k_Fl)$ are expected to capture the essential transport behavior.

To quantify this more precisely we estimate the localization length, which for a 2D system can be estimated as~\cite{RMP1985}:

\be
\xi \sim l \exp \left(\frac{1}{2} \pi  k_F l \right)
\ee

\noindent and is about $\xi \sim \{3.2 \times 10^{-5}m, 0.6 m\}$ for our Hall bar samples. This length scale needs to be compared with a density diffusion length relevant for the Altshuler-Aronov theory of interaction corrections, which, near zero bias, can be taken to be controlled by the temperature of the sample and is given by~\cite{AABook}:

\be
L_{\epsilon} \sim \sqrt{\frac{\hbar D}{k_B T }}\sim \sqrt{\frac{\hbar v_F  l}{2 k_B T}},\ D=\frac{\tau  v_F^2}{2}=\frac{l v_F}{2}
\ee

\noindent  For a typical sample temperature of T$\sim $ 50mK and using the fermi velocity of a TI with
4 QL's~\cite{zhang2010crossover} ($v_F\sim 4.5\times 10^5 m/s$), we obtain an interaction length: $L_\epsilon \sim \{ 3\times10^{-7}m,3.6\times10^{-7}m\}$. Therefore the dimensionless parameter $\log (L_{\epsilon }/l )/\log (\xi/l) \sim \{0.5,0.2\}$  is found to be reasonably small lending confidence to the applicability~\cite{RMP1985,AABook} of the Altshuler-Aronov theory to our samples. Apart from the difference in localization lengths, the order of magnitude of the estimates of transport coefficients are not very sensitive on whether we assume spinless or spin-full electrons.


\subsection{Electronic structure and range of densities for spinless vs spinful picture}

A simple model for the electronic dispersion of the two dimensional electron system arising in thin 2D Ti slabs starts for two flavors of Dirac fermions describing the states in top and bottom surfaces coupled via tunneling. This model is supplemented by a ferromagnetic exchange splitting, and is capable of predicting the quantum anomalous Hall state arising in these systems~\cite{Yu2010}. The Hamiltonian can be written as:

\begin{eqnarray}
&&H=\nonumber\\
&&\left(
\begin{array}{cccc}
 \frac{U}{2}+M_0 & i v(k_x-i k_y) & t & 0 \\
 -i v(k_x+i k_y) & \frac{U}{2}-M_0 & 0 & t \\
 t & 0 & -\frac{U}{2}+M_0 & -i v(k_x-i k_y) \\
 0 & t & i v(k_x+i k_y) & -\frac{U}{2}-M_0 \\
\end{array}
\right),\nonumber
\end{eqnarray}

\noindent where $M_0$ is the exchange splitting induced energy scale, $t$ is the tunneling strength between top and bottom surfaces, $U$ is a possible bias between top and bottom layers developed when an electric field is present across the TI slab, and $v$ is the typical Dirac velocity of the surface states of the TI. There are four bands associated with this Hamiltonian:

\small{
\be
E_k=\pm \sqrt{ v^2 k^2+M_0^2+t^2+(U/2)^2\pm\sqrt{v^2 k^2 U^2+M_0^2 \left(4 t^2+U^2\right)}}.
\ee

\noindent Assuming the field across the slab is negligible ($U\approx0$), then the dispersion reduces to:

\be
E_k=\pm \sqrt{ v^2 k^2+(|M_0|\pm |t|)^2}.
\ee}

\noindent The condition for the stability of the topologically non-trivial AQH insulator is: $|M_0| > |t|$ (the state with $|M_0| < |t|$ has a Chern number $0$ and is a trivial insulator). When the system is doped and becomes metallic, the condition the chemical potential must satisfy in order to guarantee that a single band is occupied is $|M_0|+ |t|> \mu >|M_0|-|t|$. This condition translates in terms of densities into:

\be
|n| <\frac{|M_0||t|}{\pi v^2 \hbar^2},
\ee

\noindent In other words, for densities within this range a single band is occupied and otherwise two bands are occupied. An estimate for $t\approx 0.06 eV$ for a slab of 4QL and $v = 4.5\times 10^5 m/s$~\cite{zhang2010crossover}. The density (using only $t$) $\frac{|t|^2}{\pi v^2 \hbar^2} \approx 3.3\times 10^{10} cm^{-2}$.
%

\end{document}